\documentclass{aa}
\usepackage{txfonts}
\usepackage{graphicx}
\usepackage{natbib}
\usepackage{amssymb}
\usepackage{amsfonts}
\bibpunct{(}{)}{;}{a}{}{,}

\newcommand{\Kt}{\widetilde{K}}
\newcommand{\xih}{\widehat{\xi}}
\newcommand{\Xih}{\widehat{\Xi}}
\newcommand{\Kh}{\widehat{K}}
\newcommand{\gb}{\bar{g}}
\newcommand{\gh}{\widehat{g}}
\newcommand{\Kb}{\bar{K}}
\newcommand{\xib}{\bar{\xi}}
\newcommand{\xit}{\widetilde{\xi}}
\newcommand{\xb}{\vec{x}}

\newcommand{\Ib}{\vec{I}}
\newcommand{\kappab}{\vec{\kappa}}
\newcommand{\kb}{\vec{k}}

\newcommand{\Oneb}{\vec{1}}
\newcommand{\Wc}{\mathcal{W}}

\newcommand{\gammab}{\vec{\gamma}}

\newcommand{\wb}{\vec{w}}
\newcommand{\Rb}{\mathbb{R}}
\newcommand{\zb}{\vec{z}}

\begin{document}

\title{Ly-$\alpha$ forest: efficient unbiased estimation of \\
second-order properties with missing data}

   \author{R. Vio\inst{1}
          \and
          V. D'Odorico\inst{2}
          \and
          H. Stoyan\inst{3}
          \and
          D. Stoyan\inst{3}
          }

   \offprints{R. Vio}

   \institute{Chip Computers Consulting s.r.l., Viale Don L.~Sturzo 82,
              S.Liberale di Marcon, I-30020 Venice, Italy\\
              \email{robertovio@tin.it}
         \and
             INAF - Osservatorio Astronomico di Trieste,
             via G.B. Tiepolo 11,
             I-34143 Trieste, Italy \\
              \email{dodorico@oats.inaf.it}
         \and
             Institute of Stochastics,
             TU Bergakademie Freiberg\\
             D-09596 Freiberg, Germany \\
             \email{stoyan@math.tu-freiberg.de}
             }

\date{Received .............; accepted ................}

\abstract {One important step in the statistical analysis of the
Ly-$\alpha$ forest data is the study of their {{\it second order}}
properties. Usually, this is accomplished by means of the two-point
correlation function or, alternatively, the $K$-function. In the
computation of these functions it is necessary to take into account
the presence of strong metal line complexes and strong Ly-$\alpha$ lines that can 
hidden part of the Ly-$\alpha$ forest and represent a non negligible
source of bias.} 
{In this work, we show quantitatively what are the effects of the
gaps introduced in the spectrum by the strong lines if they are not
properly accounted for in the computation of the correlation
properties. We propose a geometric method which is able to solve this problem and
is computationally more efficient than the Monte Carlo (MC) technique that is typically
adopted in Cosmology studies.  The method is implemented in two different algorithms.  
The first one permits to obtain exact results, whereas the second one
provides approximated results but is computationally very efficient.
The proposed approach can be easily extended to deal with the case of two or more lists of lines 
that have to be analyzed at the same time.} 
{Numerical experiments are presented that
illustrate the consequences to neglect the effects due to the
strong lines and the excellent performances of the proposed
approach.} {The proposed method is able to remarkably improve the
estimates of both the two-point correlation function and the
$K$-function.} {} 
\titlerunning{Unbiased estimation of the two-point correlation function}
\authorrunning{R. Vio et al.}
\maketitle

\section{Introduction}

The study of the absorption lines observed in the optical spectra of
high redshift quasars has greatly evolved in the last decade thanks
mainly to the advent of the new class of 10m-telescopes coupled with
high resolution, high performance spectrographs and to the improved
comprehension of the physical nature of the absorbers. 

The lines forming the so called ``Ly-$\alpha$ forest'' observed blue ward 
of the Ly-$\alpha$ emission of the quasar are  due to the Ly-$\alpha$
transition in neutral hydrogen distributed along the line of sight. 
They originate in the fluctuations of the intermediate and low density
intergalactic medium (IGM), arising naturally in the hierarchical
process of structure formation \citep[e.g.,][]{cen94,zhang95,
hernquist96,miralda96,bi97,dave97,zhang97,theuns98,machacek00}. 
Thus,  the Ly-$\alpha$ forest provides a unique and powerful tool to study 
the distribution/evolution of baryonic matter and the physical status of the IGM 
over a wide redshift interval.  

In analogy with the study of galaxies, the two-point correlation function 
has been classically used to study the distribution of the Ly-$\alpha$ 
lines \citep[e.g.][]{sargent80}. 
However, the Ly-$\alpha$ distribution is contaminated in the observed
spectra by metal transitions associated with galactic haloes pierced
by the line of sight. 
The stronger  metal lines can extend for several angstrom covering
weaker Ly-$\alpha$ lines and causing a bias in their distribution. The
same is true for strong Ly-$\alpha$ absorptions, in particular, Lyman
limit and damped Ly-$\alpha$ systems.  On the other hand, the
distribution at small scales is biased by the intrinsic width of the
lines ($\sim 25-30$ km s$^{-1}$). 

Observers have become more and more aware of those problems in 
particular for high resolution spectra and at large redshifts where the
crowding of the lines increases ($dn/dz \propto (1+z)^{2.5}$,
e.g., \citealp{kim02}).   
In the most recent papers based on high resolution quasar spectra
\citep[e.g.,][]{lu96,cristiani97,kim01} the strong
Ly-$\alpha$ lines are eliminated from  the computation of the two-point
correlation function and the lines closer than a minimum  
velocity separation are merged. In order to correct for edge effects
and presence of the metal gaps in the line distribution, the observed
distribution of line pairs is compared with Monte Carlo (MC)
simulations of lines which are randomly distributed in the observed
redshift interval other than the cosmological increase with
redshift. The same gaps due to the cut-out absorptions are reported
also on the simulated spectra and the minimum separation between two
lines is preserved. The simulated process is approximately of Poisson type 
over small redshift separations \citep[see][]{sargent80}. 

A further approximation is introduced in this method because it is not
the redshift split but the velocity split distribution that is
considered, where the separation $\Delta v_{ij}$ between two absorption
lines of redshifts $z_i$ and $z_j$ is given by the non linear formula
$\Delta v_{ij} = c | (z_i - z_j) | / [1+(z_i+z_j)/2]$. Since, if
$\Delta v_{13} >  \Delta v_{12}, \Delta v_{23}$, it is 
$\Delta v_{13} \neq \Delta v_{12} + \Delta v_{23}$, the quantities $\Delta v_{ij}$
do not represent Euclidean distances. Hence, from the quantities 
$\{ \Delta v_{i,j} \}_{i,j = 1}^N$ it is not even possible to construct 
the one-dimensional point process corresponding to the sequence of the velocities  
$v_1, v_2, \ldots, v_N$. 
The use of velocities at the place of comoving separations was
suggested mainly by the unknown ratio between peculiar velocity and
Hubble flow contributions to the measured Ly-$\alpha$ line redshifts.   
There is now evidence that peculiar velocities are negligible for
those absorbers \citep[e.g.,][]{rauch05,dodorico06}, which makes it
preferable to use comoving distances to compute the two-point
correlation function.

 The computation of a comoving distance
corresponding to a given $z$ requires the evaluation of an integral
that in general is not available  in closed form. Hence, a numerical
approach has to be adopted. This can make the MC methods very time
consuming. For this reason, in this paper we present a geometric  
method to compute the two-point correlation that does not require the
generation of auxiliary random  samples and therefore is much more
efficient.  

In Sec.~\ref{sec:edge} the computation of the two-point correlation function
and of the $K$-function is considered when only edge effects are present.
In Sec.~\ref{sec:gaps} and Appendices~\ref{sec:helga}-\ref{sec:functions} the arguments are generalized
to deal with the case of gaps in the data sequence and two algorithms are proposed.
Two experimental data sets are analyzed in Sec.~\ref{sec:application}. Finally,
conclusions are presented in Sec.~\ref{sec:conclusions}.

Throughout this paper we will adopt for the cosmological parameters: 
$\Omega_{0\rm{m}} = 0.26,\ \Omega_{0\Lambda}=0.74$, and  $H_0=73$ km
s$^{-1}$ Mpc$^{-1}$. 

\section{Estimation of $K(r)$ and $\xi(r)$ in presence of edge
  effects} \label{sec:edge} 

A one-dimensional spatial point process $X$ is any stochastic
mechanism that generates a countable set of events $\xb = \{ x_i
\}_{i=1}^n$ on the real axis $\Rb$. It is called {\it stationary} if
the distribution of $X$ is invariant under translations, that is,
the distribution of $X + s$ is the same as that of $X$ for any $s
\in \Rb$. Usually, stationary point processes are described by their
{\it first-order} and {\it second-order} characteristics. The {\it
first-order} characteristic, say $\lambda$, equals the mean number
of points per unit of length and is often called {\it intensity}.
The second-order characteristics are represented by the two-point
correlation function $\xi(r)$ and the pair correlation function
$g(r)$ that satisfy
\begin{equation}
g(r) = 1 + \xi(r).
\end{equation}
The function $g(r)$ is defined as follows. Consider any
infinitesimal interval $B$ of length $dL$. The probability of having
a point in $B$ is $\lambda dL$. Consider now two such intervals
$B_1$ and $B_2$ of lengths $dL_1$ and $dL_2$ and intercenter
distance $r$. The probability to have a point in each interval can
be denoted by $P(r)$ and is expressed as
\begin{equation}
P(r) = g(r) \lambda^2 dL_1 dL_2.
\end{equation}
The factor of proportionality $g(r)$ is the pair correlation
function. It is clear that, in the case of complete randomness,
$g(r)=1$ or $\xi(r) = 0$, independently of $r$. $g(r) > 1$, or
$\xi(r) > 0$, indicates that pairs of points with distance in the
interval $[r-dr/2, r+dr/2]$ are more likely to occur than for a
Poisson process with the same intensity, that means there is
clustering. On the contrary,  $g(r) < 1$ or $\xi(r) < 0$ correspond to
inhibition or regularity.

The estimation of $g(r)$ and $\xi(r)$ presents a practical
difficulty. In fact, for a given distance $r$, the estimation of
these characteristics requires evaluation of the number of point
pairs in the infinitesimal interval $[r-dr/2, r+dr/2]$. Since this
is impossible, a finite interval $[r - h, r + h]$ has to be used.
This leads to approximation errors, especially for small $r$ when it
could happen that $r - h < 0$, and to the problem of choosing the
bandwidth $h$. In particular, the determination of the {\it best}
value of $h$ to use for a specific problem is still an "art"
\citep{mar05}. The best recipe is to
consider a series of bandwidths starting from $h\approx 1/\lambda$
and then to compare the results; sometimes it is
recommendable to use ``adapted'' bandwidths, i.e. different values
of $h$ for small and large $r$. In order to avoid this kind of
problem, sometimes statisticians use also the so-called
$K$-function, $K(r)$, that is related to $\xi(r)$ through
\begin{equation} \label{eq:k}
K(r) = 2 \int_0^r [1 + \xi(u)] du,
\end{equation}
or
\begin{equation} \label{eq:xi}
\xi(r) = \frac{1}{2} \frac{dK(r)}{dr} -1.
\end{equation}
Here, no bandwidth has to be fixed.
The main drawback of this approach is that $K(r)$ is a cumulative function, which increases
nearly linearly. (A comparable situation exists in classical
statistics for the pair ``probability distribution function'' $F(x)$
and the ``probability density function'' $f(x)$.) For this reason, a plot of $K(r)$
is more difficult to interpret than a plot of $\xi(r)$. As a
consequence, $\xi(r)$ is more useful in exploratory statistics,
whereas $K(r)$ is useful in testing statistical hypotheses, e.g.,
that a given point sequence is ``completely random'' or belongs to a Poisson
process.

If a point process was observable over the entire $\Rb$ domain, the
estimation of $K(r)$ and $\xi(r)$ could simply follow their
definitions. But in general, a point process can be observed only in
a finite region $W$ defined in the interval $[x_{\rm min}, x_{\rm max}] \subset \Rb$,
the {\it observing window}. Then {\it edge
effects} play an important role, since the estimation of $K(r)$ and
$\xi(r)$ should require the use of points external to $W$. Hence,
the ``{\it natural}'' estimator
\begin{equation} \label{eq:kn}
\Kb(r) = \frac{|W|}{n^2} \sum_{i=1}^n \sum_{j =1, j \neq i}^n I_r(|x_i - x_j|),
\end{equation}
with $|W|$ the length of $W$ and
\begin{equation}
I_r(u) = \left\{
\begin{array}{ll}
1 & \textrm{~~if $u \leq r$} \\
0 & \textrm{~~otherwise,}
\end{array}
\right.
\end{equation}
will yield results too small, with a negative bias.
The same holds for the ``{\it natural}'' estimator of $\xi(r)$
\begin{equation}
\xib(r) = \frac{|W|^2}{2 n^2} \sum_{i=1}^n \sum_{j =1, j \neq
i}^n k_h(r-|x_i - x_j|) - 1,
\end{equation}
where
\begin{equation}
k_h(u) = \left\{
\begin{array}{ll}
1/2 h & \textrm{~~if $- h \leq u \leq h$} \\
0 & \textrm{~~otherwise.}
\end{array}
\right.
\end{equation}

\begin{figure}
        \resizebox{\hsize}{!}{\includegraphics{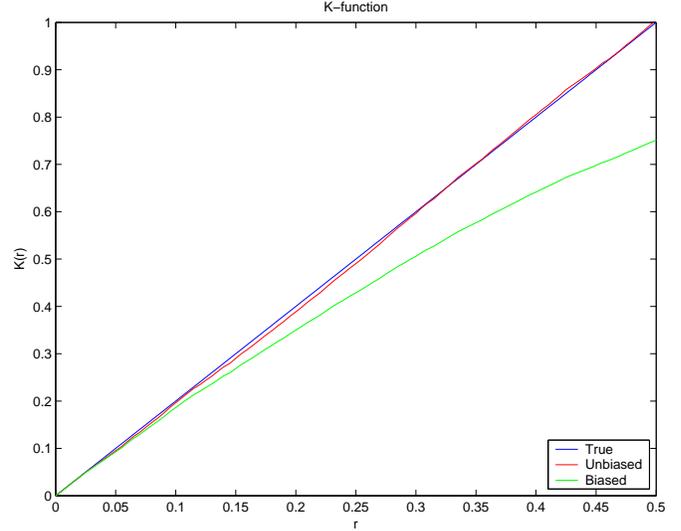}}
        \caption{Comparison of the biased estimator $\Kb(r)$ with the
        unbiased one $\Kh(r)$ with adapted intensity $\lambda_V(r)$. The data used in the simulation
        consist of $200$ points drawn from a Poisson process in the interval $[0, 1]$.}
\label{fig:fig_kcomp}
\end{figure}

\begin{figure}
        \resizebox{\hsize}{!}{\includegraphics{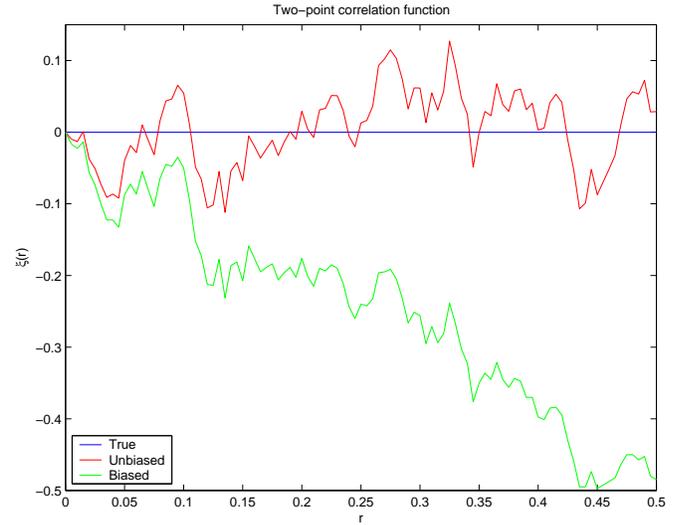}}
        \caption{Comparison of the biased estimator $\gb(r)$ with the
        unbiased one $\gh(r)$ with adapted intensity $\lambda_S(r)$.  The data used in the simulation are
      the same of Fig.~\ref{fig:fig_kcomp}. $h=0.05$.}
\label{fig:fig_gcomp}
\end{figure}

Figures~\ref{fig:fig_kcomp}-\ref{fig:fig_gcomp} show that the
mentioned biases are really significant also in the one-dimensional
case considered here. As expected, the larger $r$ the larger the
biases. This is due to the increasing number of
points external to $W$ that, for increasing values of $r$, should be
necessary to estimate both $K(r)$ and $\xi(r)$. Although not
clearly visible in Fig.~\ref{fig:fig_gcomp}, $\xib(r)$ presents a
bias also when $r < h$. This is because in the computation of
$\xi(r)$ only the points in the interval
$[0, r + h]$ of length shorter than $2 h$ can be considered.

In Astronomy, especially in the field of the statistical analysis of
the systems of galaxies, this edge-correction problem is well known. 
Various algorithms, mostly of Monte Carlo (MC) type, have been proposed for
the three-dimensional case \citep[e.g., see ][ and references
therein]{ham93, lan93, qua99, pon99, ker99, ker00, mar02}. Their
conversion to the one-dimensional case is straightforward. However, the fact to work in 
the one-dimensional case allows us to follow the approach due to \citet{sto00} which is a
modification of the MC methods presented in \citet{ham93} and \citet{lan93}.
In practice, the auxiliary random samples of points used in
those papers are replaced by geometrical terms that result from exact integration.
This permits to obtain more precise results since
the  quantities of interest are {\it calculated} and not only
{\it estimated} as in the case of the MC techniques. Moreover, there
is a remarkable benefit in the computational efficiency. In
fact, in order to provide satisfactory results, the MC methods
require the use of a number of auxiliary random samples much larger
than the number of experimental points with obvious computational
costs \citep[see][]{mar02}.

We start from the unbiased {\it rigid motion corrected} estimators of
$\lambda^2 K(r)$ and $\lambda^2 \xi(r)$:
\begin{equation} \label{eq:k1}
\lambda^2\Kh(r) = |W| \sum_{i=1}^n \sum_{j =1, j \neq i}^n
p_{ij}^{-1} I_r(|x_i - x_j|),
\end{equation}
where the weights $p_{ij}$ are given by
\begin{equation} \label{eq:w}
p_{ij} = \frac{| W \cap W_{x_i-x_j} |}{|W|}.
\end{equation}
$W_s$ is the {\it observing window} shifted by the amount $s$.
If $W$ is an interval, it is simply $|W \cap W_{x_i-x_j}| = |W| -
|x_i - x_j|$. Similarly:
\begin{equation} \label{eq:xi1}
\lambda^2\xih(r) =  \frac{|W| c(r)}{2}
\sum_{i=1}^n \sum_{j=1, j \neq i}^n p_{ij}^{-1} k_h(r - |x_i - x_j|) - \lambda^2.
\end{equation}
Here, $c(r) = 2 h /( r + h - \max[0, r - h])$ is a correcting factor
for the values of $r$ for which $r - h$ goes negative \citep{gua06}. A ``naive''
method used by many authors to obtain estimates of $K(r)$ and $\xi(r)$ 
is to set $\lambda^2 = n^2/|W|^2$ and then to divide the 
estimators~(\ref{eq:k1}) and
(\ref{eq:xi1}) by this quantity. However, the circumstance that
$n/|W|$ is a good estimator of $\lambda$ does not necessarily
implies that the same holds for $n^2/|W|^2$ with respect to
$\lambda^2$. In fact, more is possible. As \citet{ham93} and \citet{lan93}
showed for $\xi(r)$ and \citet{sto00} for $K(r)$, it is possible to
reduce the {\it mean squared error} (${\rm MSE}$) \footnote{The {\it
mean squared error} of an estimator $\xit(x)$ of $\xi(r)$ is given
by ${\rm MSE}_{\xit}(r)={\rm E}[(\xit(r) - \xi(r))^2]$. A similar
expression holds for ${\rm MSE}_{\Kt}(r)$.} if, instead of the naive
classical estimator, adapted estimators for $\lambda$ are used. One
particularity of these estimators is that they are different for
$K(r)$ and $\xi(r)$ and depend on the distance $r$. They are denoted
by $\lambda_S(r)$, for $\xi(r)$, and $\lambda_V(r)$, for $K(r)$. The
index ``$S$'' refers to {\it surface} and ``$V$'' to volume
\citep[for more details, see][]{sto00}. The expression for the two 
estimators is given by
\begin{equation} \label{eq:ls}
\lambda_S(r) = \sum_{i=1}^n \frac{ \Oneb_W(x_i-r)+\Oneb_W(x_i+r)}{2  |W \cap W_r| },
\end{equation}
and
\begin{equation} \label{eq:lv}
\lambda_V(r) = \sum_{i=1}^n \frac{ | W \cap [x_i-r, x_i+r] |}{2 \int_0^r  |W \cap W_t| dt},
\end{equation}
with
\begin{equation}
\Oneb_w(x) = \left\{
\begin{array}{ll}
1 & \textrm{~~if $x \in W $}, \\
0 & \textrm{~~otherwise.}
\end{array}
\right.
\end{equation}
Figures~\ref{fig:fig_kcomp}-\ref{fig:fig_gcomp} show the remarkable improvement
in the estimate of both $K(r)$ and $\xi(r)$.

\section{Estimation of $K(r)$ and $\xi(r)$ in presence of gaps} \label{sec:gaps}

\begin{figure}
        \resizebox{\hsize}{!}{\includegraphics{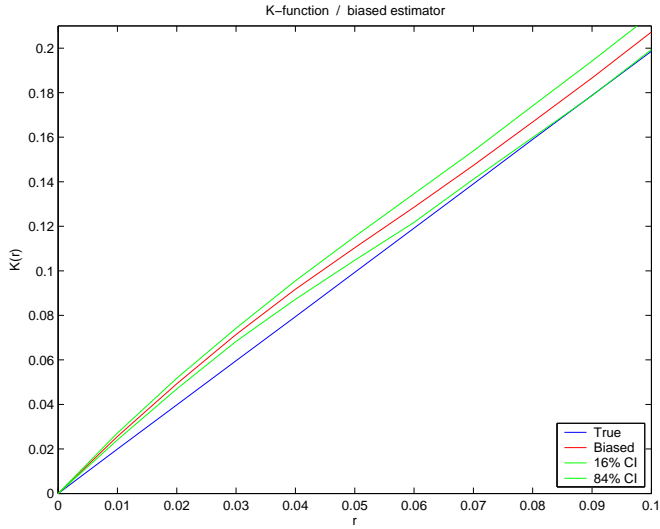}}
        \caption{Mean and $68\%$ (i.e., $1$-$\sigma$) {\it confidence interval} for the
      biased estimator $\Kb(r)$, with periodic 
      boundary conditions, obtained from $100$ realizations of
        a Poisson process with gaps. The data sets has been simulated
      by drawing $200$ points from a Poisson process in 
        the interval $[0, 1]$ and discharging those that are in subset
      $[0.20, 0.25] \cup [0.30, 0.35] \cup [0.50, 0.55] \cup [0.70,
      0.75] \cup [0.80, 0.85]$. For reference also the $K$-function of 
        the Poisson process is shown (``true'' line).
         Because of the imposed period boundary conditions, estimator
      $\Kb(r)$ is free from the edge effects.} 
        \label{fig:fig_kbi_po}
\end{figure}

\begin{figure}
        \resizebox{\hsize}{!}{\includegraphics{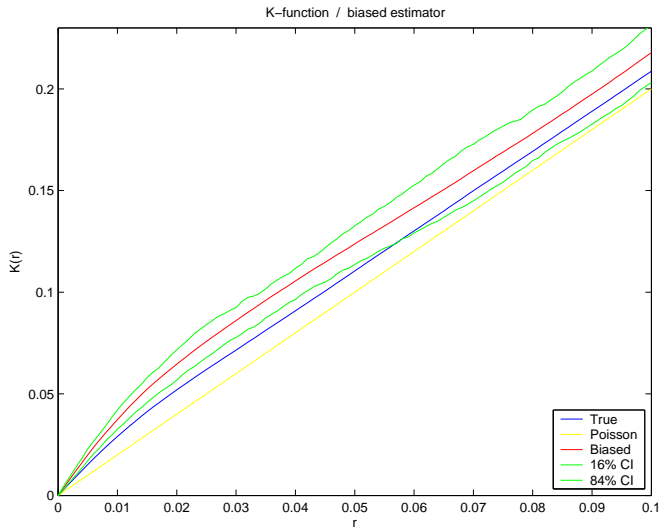}}
        \caption{The same as in Fig.~\ref{fig:fig_kbi_po} but for a
        Cox process simulated in the interval $[0, 1]$ 
        by taking the absolute value of a stationary Gaussian random
        field with a Gaussian correlation function whose dispersion is
        set to $0.01$. The Gaussian random fields have been generated 
        under the hypothesis of periodic boundary conditions. Because of this, the
        line labeled ``{\it True}'', that provides the mean value of
        the estimator $\Kb(r)$ with periodic boundary conditions, is
        free from the edge effects. This mean has been computed on the
        basis of $5000$ simulations with $W \equiv [0, 1]$. For
        reference, also the $K$-function of a Poisson process is
        shown.} 
        \label{fig:fig_kbi_ns}
\end{figure}

In practical applications, the edge effects are not the only
problem. In fact, often the observing window
is a union of intervals separated by gaps. 
In the spectra of quasars this is due to the presence of 
metal lines and strong Ly-$\alpha$ systems  that can hidden portions 
of the Ly-$\alpha$ forest. 
Ignoring these gaps and taking 
erroneously the interval starting with the most left interval
start-point to the most right interval endpoint as the observing window
produces biased estimates. Clearly, the estimators will tend to
report more clustering than really existing. The reason is simply
that when the points within some given regions are removed, then
this artificially creates separated groups of points that mimic the
presence of clusters. As a consequence, an erroneous indication of
clustering can happen even in the case of {\it no-interaction} or
{\it complete spatial randomness} in the point pattern. On the other
hand, if clustering or aggregation is really present, a
higher degree of clustering than true may be obtained. These effects
are clearly visible in
Figs.~\ref{fig:fig_kbi_po}-\ref{fig:fig_kbi_ns} where the
performance of the estimator $\Kb(r)$ is shown for a Poisson
process, typically of complete random spatial patterns, and a Cox
one that models spatial patterns with significant density fluctuations
\citep{mol04}.

In presence of gaps the unbiased estimators $\Kh(r)$ and $\xih(r)$
and the intensities $\lambda_V(r)$ and $\lambda_S(r)$ can still be
used. Here, however, a computational problem arises. In fact,
although the computation of the quantities $|W \cap W_{x_i-x_j}|$,
$\lambda_S(r)$ and $\lambda_V(r)$ needs only elementary mathematics,
nevertheless it is impossible to give explicit formulas. For small
$r$, a simple approximation is given by
\begin{equation} |W \cap W_r| = L - (n_w+1)r,
\end{equation}
where $L$ is the distance between the most left observation interval
endpoint and the most right observation interval endpoint and $n_w$
is the number of gaps. For larger $r$, the exact algorithm
presented in Appendix~\ref{sec:helga} can be used. As an alternative, the approximated but
efficient algorithm  based on a Fourier technique
in Appendix~\ref{sec:approximated} is proposed. After
that, it is possible to compute $\lambda_S(V)$ and then $\xih(r)$.
The computation of $\lambda_V(r)$ is more complicated and can be
carried out by numerical integration. However a more efficient approach, 
based again on a Fourier technique, is described in 
Appendices~\ref{sec:approximated}-\ref{sec:lambdas}.

\begin{figure}
        \resizebox{\hsize}{!}{\includegraphics{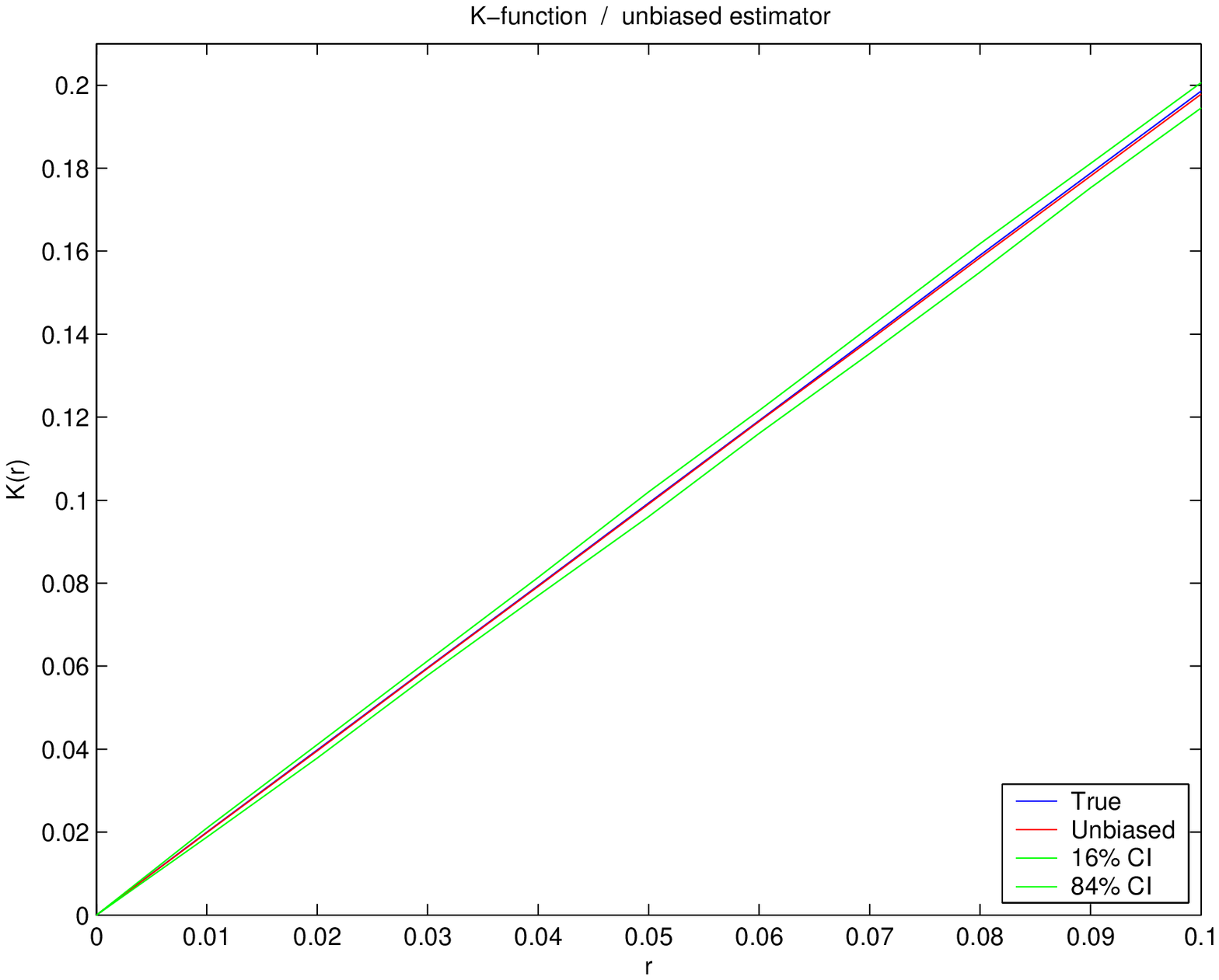}}
        \caption{As in Fig.~\ref{fig:fig_kbi_po} but for the unbiased
        estimator $\Kh(r)$ and the adapted intensity $\lambda_V(r)$.}
        \label{fig:fig_kun_po}
\end{figure}

\begin{figure}
        \resizebox{\hsize}{!}{\includegraphics{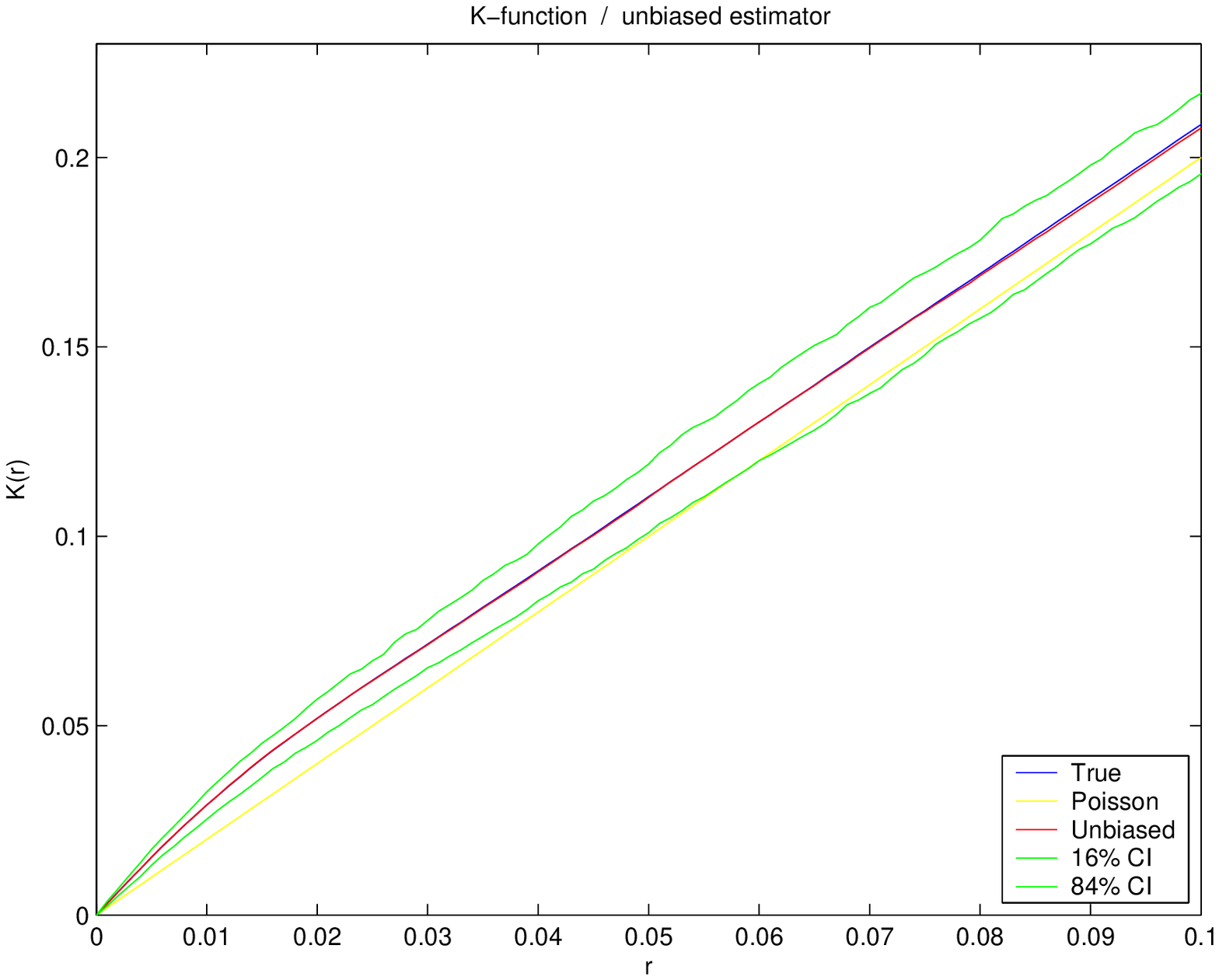}}
        \caption{As in Fig.~\ref{fig:fig_kbi_ns} but for the unbiased
        estimator $\Kh(r)$ and the adapted intensity $\lambda_V(r)$.}
        \label{fig:fig_kun_ns}
\end{figure}

\begin{figure}
        \resizebox{\hsize}{!}{\includegraphics{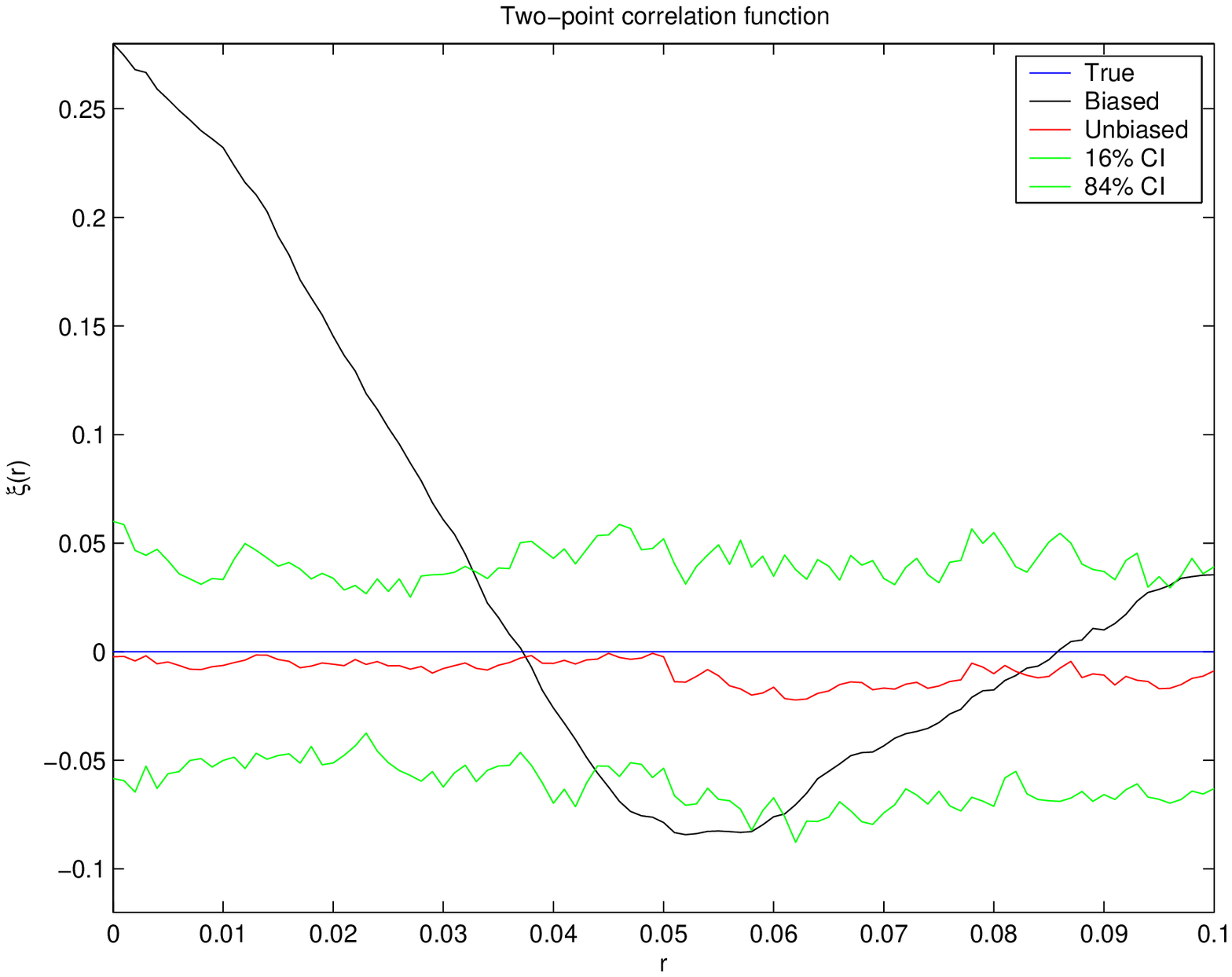}}
        \caption{As in Fig.~\ref{fig:fig_kbi_po} but for the unbiased
        estimator $\xih(r)$ and the adapted intensity $\lambda_S(r)$. $h=0.01$.}
        \label{fig:fig_corr_po}
\end{figure}

\begin{figure}
        \resizebox{\hsize}{!}{\includegraphics{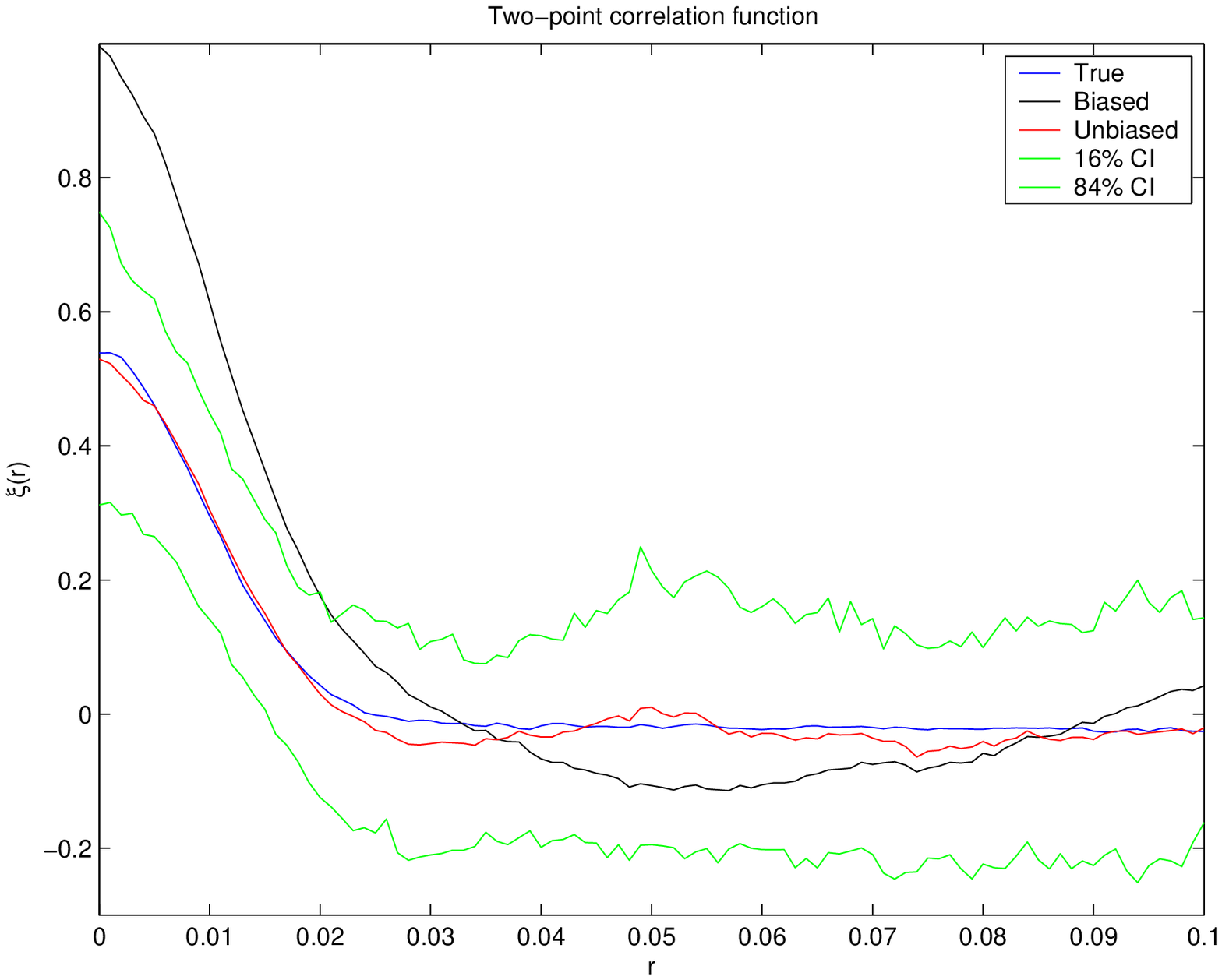}}
        \caption{As in Fig.~\ref{fig:fig_kbi_ns} but for the unbiased
        estimator $\xih(r)$ and the adapted intensity
        $\lambda_S(r)$. $h=0.005$.} 
        \label{fig:fig_corr_ns}
\end{figure}

\begin{figure}
        \resizebox{\hsize}{!}{\includegraphics{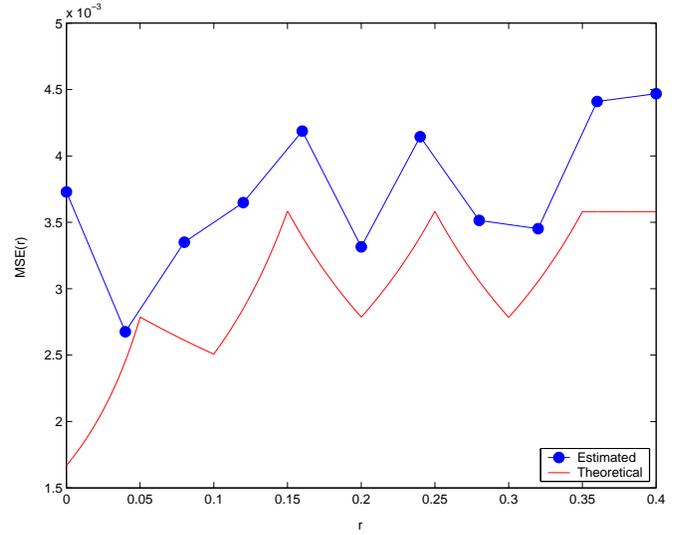}}
        \caption{Estimated ${\rm MSE}_{\xi}(r)$ (calculated for eleven values of $r$)
        vs. the theoretical one for the experiment in Fig.~\ref{fig:fig_corr_po}. 
        Notice the non-monotonic behavior of the two functions that is 
        directly linked to the fact that the {\it observing window} $W$ is constituted by disjoint 
        segments.}
        \label{fig:mse}
\end{figure}

The excellent performance of the estimator $\Kh(r)$ combined
with the intensity $\lambda_V(r)$ is shown in
Figs.~\ref{fig:fig_kun_po}-\ref{fig:fig_kun_ns} (to compare with
Figs.~\ref{fig:fig_kbi_po}-\ref{fig:fig_kbi_ns}).
Figures.~\ref{fig:fig_corr_po}-\ref{fig:fig_corr_ns} show the
results obtainable with $\xih(r)$ combined with $\lambda_S(r)$. The
number of points used in this simulations is of the same order of
that typically expected for the Ly-$\alpha$ lines in a single spectrum. In
order to check if these results are consistent with what expected
from the theory, in Fig.~\ref{fig:mse} the theoretical ${\rm
MSE}_{\xi}(r)$ of the $\xih(r)$ estimator combined with
$\lambda_S(r)$ for the experiment with the Poisson process used
for Fig.~\ref{fig:fig_corr_po} is compared with that estimated
on the basis of $5000$ simulations; apart from the very small values
of $r$, where the effects due to the bandwidth $h$ are more
important, the agreement is reasonably good. Worth of note is the non-monotonic 
behavior of the two ${\rm MSE}$s that is directly linked to the fact that
the {\it observing window} $W$ is constituted by disjoint segments.
The equation used for ${\rm MSE}_{\xi}(r)$,
\begin{equation}
{\rm MSE}_{\xi}(r) = \frac{\xi(r)+1}{2 h \lambda^2 |W\cap W_r|},
\end{equation}
is due to \citet{sto93}. 

\section{A practical application} \label{sec:application}

In order to apply the discussed method and verify its reliability, we
have used the best high resolution high signal-to-noise quasar spectra
publicly available observed with the UVES spectrograph \citep{dekker}
at the VLT telescope in the context of the ESO Large Programme ``The
Cosmic Evolution of the IGM'' \citep[][]{bergeron04}.    
In particular, we have chosen two examples of Ly-$\alpha$ forest
contaminated by a large number of metal transitions: PKS0237-23
at $z_{\rm em}=2.233$ and PKS2126-158 at $z_{\rm em}=3.267$, for which
the line blanketing due to the same Ly-$\alpha$ lines is more severe
as the emission redshift is larger. 

The quasar spectra have a resolution $R \simeq 45000$ and a
signal-to-noise ratio in the Ly-$\alpha$ forests $S/N\simeq 80$ per
pixel. They have been reduced with the pipeline  of the instrument
\citep[version 2.1, ][]{ball00} provided by ESO in the context of the
data reduction package MIDAS.   
The continuum level was determined with a manual subjective method
based on the selection of the regions free from clear absorption that
are successively fitted with a spline polynomial of 3rd degree.  
The normalized spectra are then inspected to identify metal absorption
systems with a particular attention to the transitions that fall in
the Ly-$\alpha$ forest. Once the forest has been cleaned, the \ion{H}{i} lines
are fitted with Voigt profiles in the LYMAN context of the MIDAS
reduction package. In order to increase the reliability of the fit
parameters for the saturated Ly-$\alpha$ lines we used also the other
unblended lines of the Lyman series falling in the observed spectral
range. We excluded 1000 km s$^{-1}$ from the Ly-$\beta$ emission and 5000 km
s$^{-1}$ from the Ly-$\alpha$ emission of the quasar to avoid
associated absorbers and the proximity effect due to the quasar
itself. Furthermore, we cut out of the spectra the regions  occupied
by the metal transitions and the Lyman limit systems (Ly-$\alpha$ lines with column 
density $N($HI$) \ge 10^{17}$ cm$^{-2}$) which could  
mask the presence of weaker lines in their velocity profiles (line blanketing). 
At the smallest scales the clustering properties are affected by the typical velocity 
width of the lines ($\sim 25-30$ km s$^{-1}$,  e.g. Kim et al. 2001), for this reason we 
have merged the line pairs closer than $25$ km s$^{-1}$.

\begin{figure}
        \resizebox{\hsize}{!}{\includegraphics{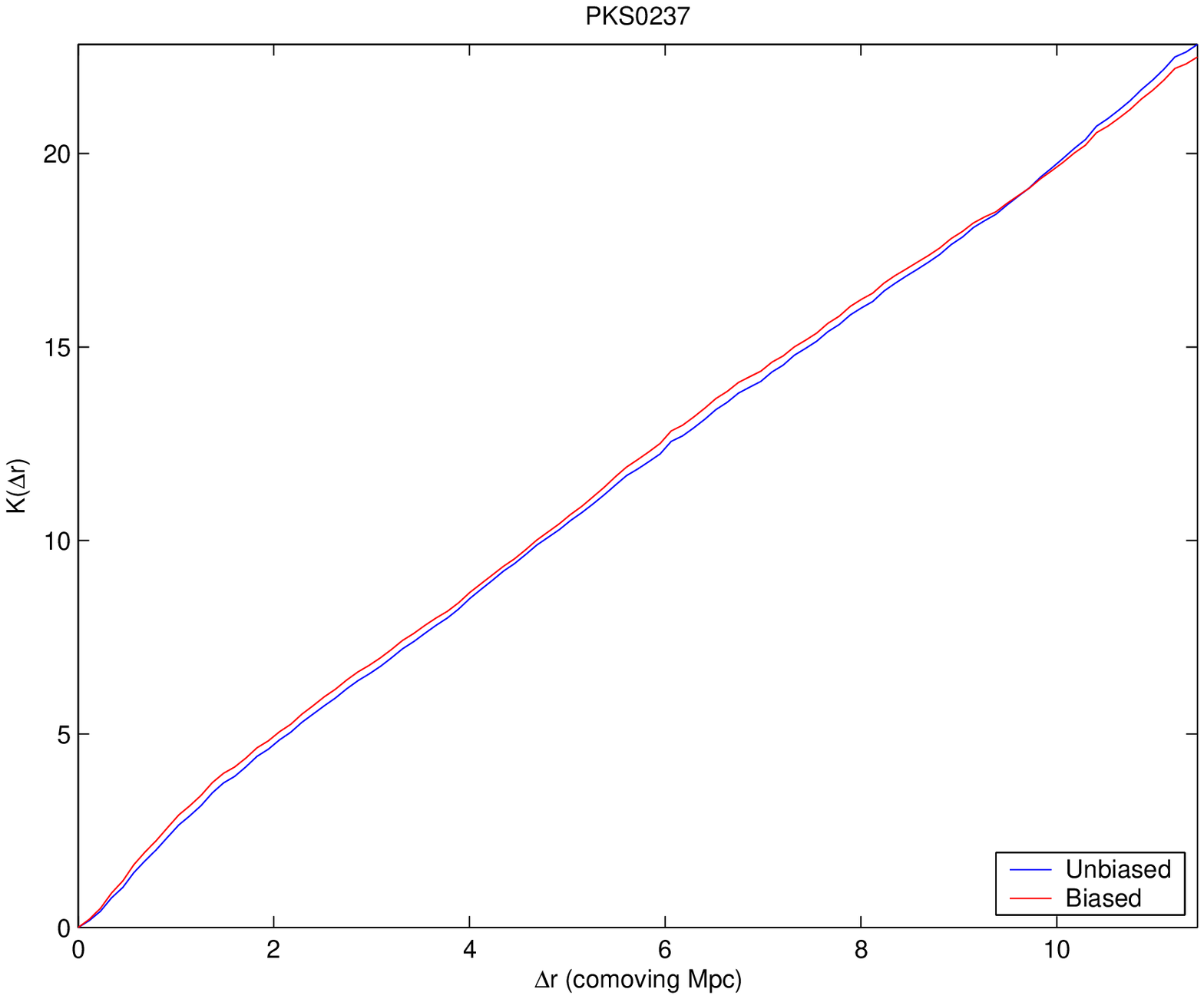}}
        \caption{$\Kh(\Delta r)$ for the data of PKS0237-23 when the gaps
        due to the metal and the strong Ly-$\alpha$ lines are taken into
        account (unbiased estimator) as a function of the comoving
        spatial separation, $\Delta r$.  The largest value of $\Delta
        r$ in the figure corresponds to about $5\%$ the interval spanned by this
        quantity. The total extension of gaps amount to about $16\%$
        of spatial interval covered by $W$. For reference, the same
        estimate is shown when gaps are neglected (biased estimator).}
        \label{fig:K_pks0237}
\end{figure}

\begin{figure}
        \resizebox{\hsize}{!}{\includegraphics{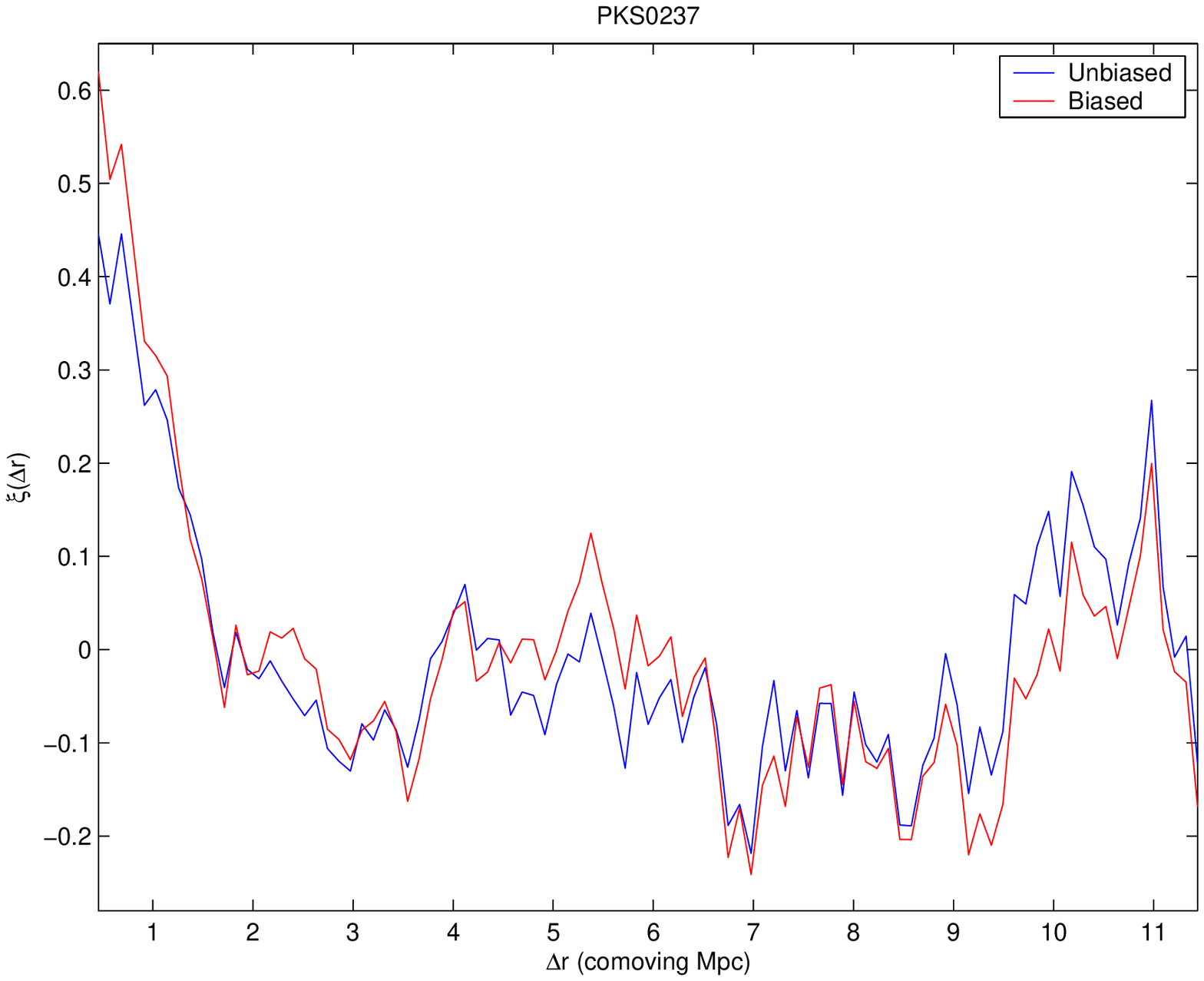}}
        \caption{$\xih(\Delta r)$ for the data of PKS0237-23 when the gaps
        due to the metal and the strong Ly-$\alpha$ lines are taken into
        account (unbiased estimator) as a function of the comoving
        spatial separation, $\Delta r$. The largest value of $\Delta
        r$ in the figure corresponds to about $5\%$ the interval spanned by this
        quantity. The total extension of gaps
        amount to about $16\%$ of spatial interval 
        covered by $W$ and $h=0.001$ in the same units. For reference,
        the same estimate is shown 
        when gaps are neglected (biased estimator).}
        \label{fig:C_pks0237}
\end{figure}

\begin{figure}
        \resizebox{\hsize}{!}{\includegraphics{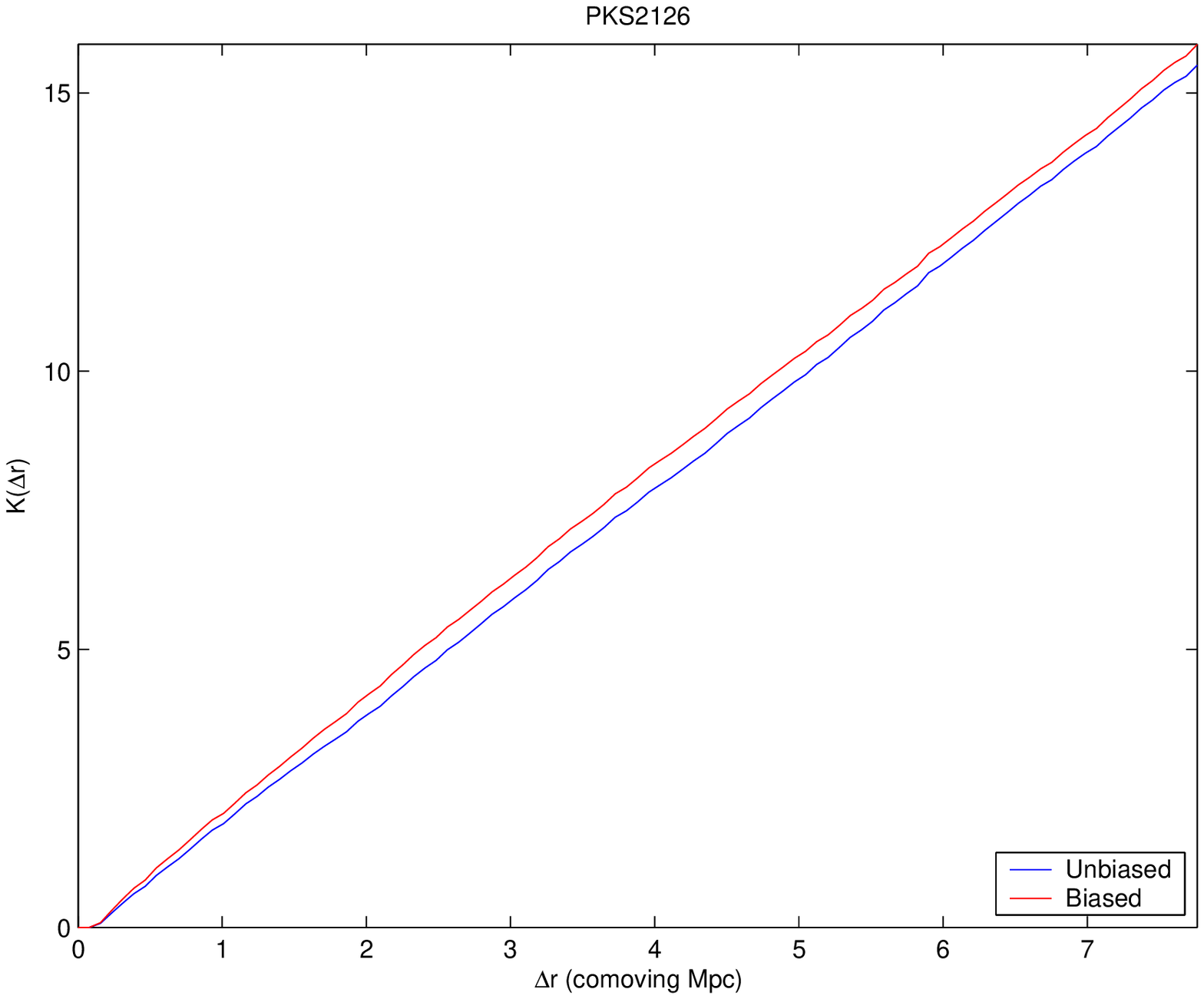}}
        \caption{$\Kh(\Delta r)$ for the data of PKS2126-158 when the gaps
        due to the metal and the strong Ly-$\alpha$ lines are taken into
        account (unbiased estimator)as a function of the comoving
        spatial separation, $\Delta r$. The largest value of $\Delta r$
        in the figure corresponds to about $5\%$ the interval spanned
        by this quantity. The total extension of gaps amount to 
        about $20\%$ of spatial interval covered by $W$. For
        reference, the same estimator is shown when gaps are 
        neglected (biased estimator).}
        \label{fig:K_pks2126}
\end{figure}

\begin{figure}
        \resizebox{\hsize}{!}{\includegraphics{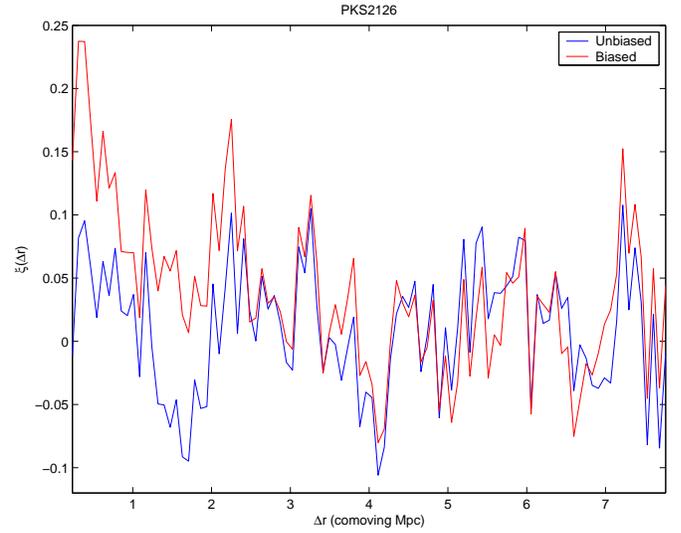}}
        \caption{$\xih(\Delta r)$ for the data of PKS2126-158 when the gaps
        due to the metal and the strong Ly-$\alpha$ lines are taken
        into account (unbiased estimator) as a function of the comoving
        spatial separation, $\Delta r$. The total extension of gaps amount to
        about $20\%$ of spatial interval covered by $W$ and $h=0.001$
        in the same units. For reference, the same estimator is shown 
        when gaps are neglected (biased estimator).}
        \label{fig:C_pks2126}
\end{figure}

\begin{figure}
        \resizebox{\hsize}{!}{\includegraphics{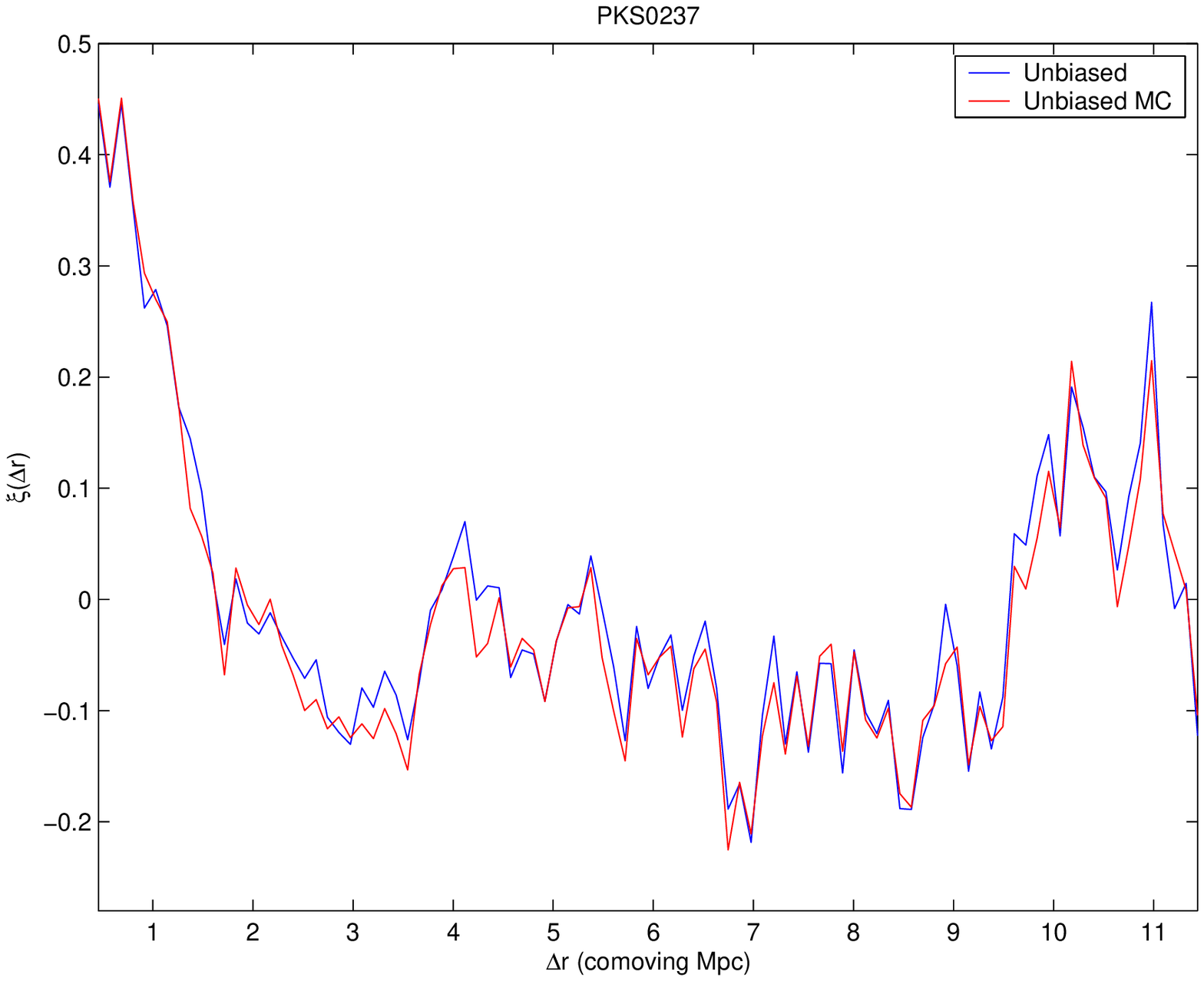}}
        \caption{$\xih(\Delta r)$ for the data of PKS0237-23 when the gaps
        due to the metal and the strong Ly-$\alpha$ lines are taken into
        account (unbiased estimator) as a function of the comoving
        spatial separation, $\Delta r$. The largest value of $\Delta
        r$ in the figure corresponds to about $5\%$ the interval spanned by this
        quantity. The total extension of gaps amount to about $16\%$ of spatial interval 
        covered by $W$ and $h=0.001$ in the same units. For reference,
        the result provided by the MC method by \cite{lan93} is also shown
        (unbiased MC estimator).}
        \label{fig:C_pks0237Com}
\end{figure}

\begin{figure}
        \resizebox{\hsize}{!}{\includegraphics{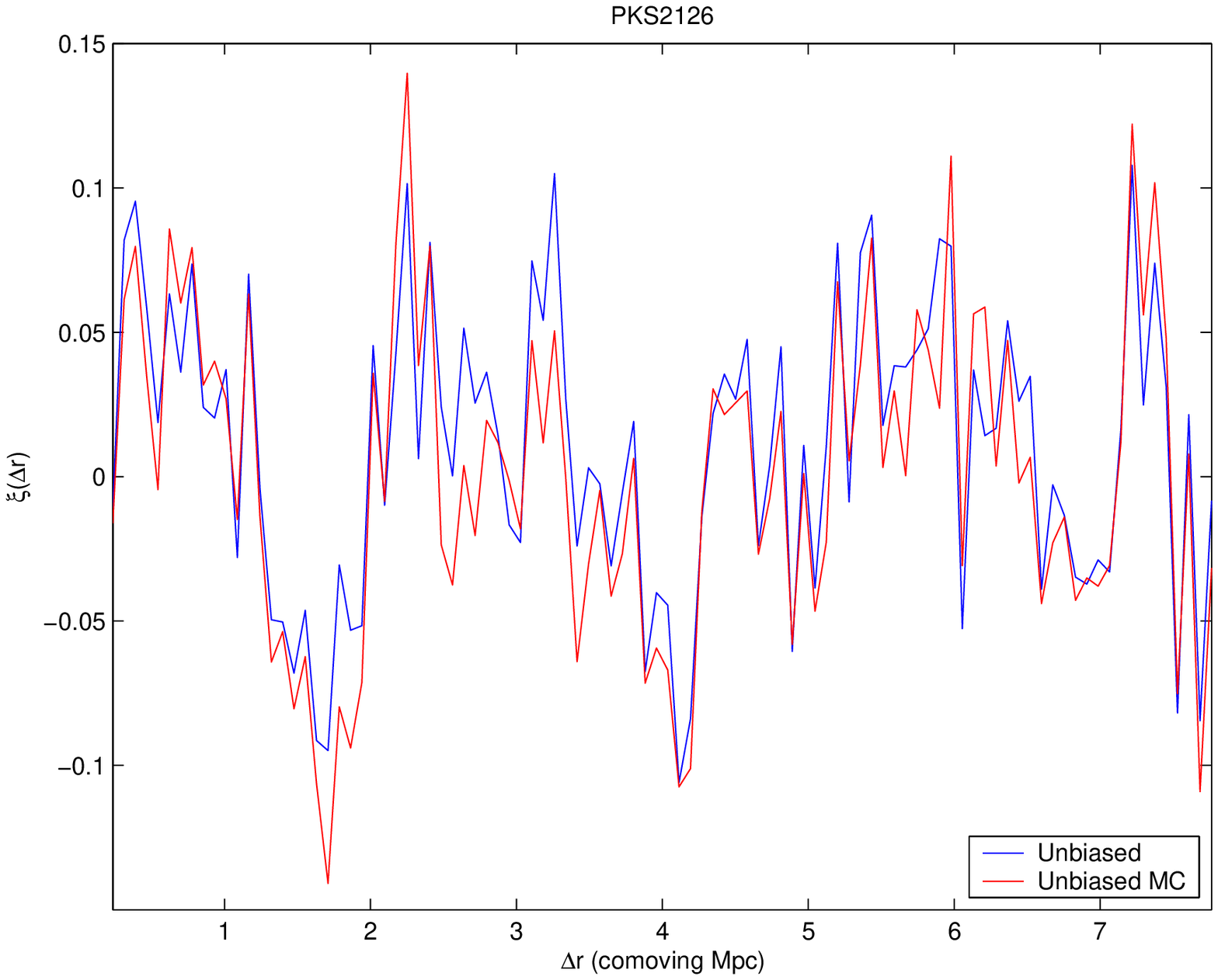}}
        \caption{$\xih(\Delta r)$ for the data of PKS2126-158 when the gaps
        due to the metal and the strong Ly-$\alpha$ lines are taken into
        account (unbiased estimator) as a function of the comoving
        spatial separation, $\Delta r$. The largest value of $\Delta
        r$ in the figure corresponds to about $5\%$ the interval spanned by this
        quantity. The total extension of gaps amount to about $16\%$ of spatial interval 
        covered by $W$ and $h=0.001$ in the same units. For reference,
        the result provided by the MC method by \cite{lan93} is also shown
        (unbiased MC estimator).}
        \label{fig:C_pks2126Com}
\end{figure}

Figures~\ref{fig:K_pks0237}-\ref{fig:C_pks2126} show $\Kh(\Delta r)$ and
$\xih(\Delta r)$  as a function of comoving spatial separation, $\Delta r$, for the
Ly-$\alpha$ forest data of the objects PKS0237-23 and PKS2126-158.  
For PKS0237-23 the number of Ly-$\alpha$ lines is $191$ and the total
length of gaps amounts to about $16 \%$ of the interval spanned by
$W$, whereas for PKS2126-158 these quantities are $300$ and $20 \%$,
respectively. For reference, the corresponding functions
edge-corrected but evaluated without considering the gaps are also
displayed. In both the cases, it is evident the
stronger indication of clustering if the gaps are not taken into
account. For the same experiment, Figs.~\ref{fig:C_pks0237Com}-\ref{fig:C_pks2126Com} show the comparison
between $\xih(\Delta r)$ and the estimate of the two-point correlation function
obtained with the MC method by \citet{lan93}. In this last case a number of $N_D=50$ data sets
have been generated that contain a number of random samples equal to
that of the original ones. $N_D$ has been chosen large enough that
the mean of the estimated two-point correlation functions does not change significantly when more data sets are
added. The good agreement is evident. However, if the 
computation of $\xih(\Delta r)$ takes a total of
$1$-$2$ seconds of CPU time \footnote{Experiments have been conducted with Matlab 6
on a {\it Pentium IV} - $1500~{\rm GHz}$ processor in a {\it Windows 2000} operating system.},
the MC method requires a similar amount of time
for each random sample. Now, if one takes into account that usually the bandwidth 
$h$ has to be determined by a {\it trail and error} approach, also with the limited size
of the data sets here used, the computational advantage of the method presented in this paper 
is obvious. This is especially true if two or more lists of lines have to be analyzed at the same time
(see below).

\section{Multi-list analysis}

Here, we briefly illustrate an additional benefit in using the estimator $\xih(r)$. In particular,
the fact that it can be easily generalized for the estimation of a
two-point correlation function when two or more lists of lines are available.
The main problem is that, in an experimental context, the lists can have different characteristics concerning the
covered spatial intervals and/or the total extension of the gaps. Hence, the naive estimator
\begin{equation}
\Xi(r) = \frac{1}{L} \sum_{l=1}^{L} \xih_l(r),
\end{equation}
with $\xih_l(r)$ the estimate obtained from the $l$th data set,
can provide unsatisfactory results since all the lists have the same importance in forming the final result. 
Some kind of weighting is necessary. In this respect, it has been proved \citep{osh00} that the estimator
\begin{equation}
\Xih(r) = \frac{\sum_{l=1}^{L} |W \cap W_r|_l ~\xih_l(r)}{\sum_{l=1}^L |W \cap W_r|_l}
\end{equation}
provides an estimate of $\xi(r)$ with a smaller variance with respect to $\Xi(r)$ for each value of $r$.
With the numerical techniques described in Sects.~\ref{sec:approximated}-\ref{sec:functions}, the implementation
of the algorithm for the computation of $\Xih(r)$ is trivial. However, the characteristics and the performances of this 
estimator are beyond the scope of the present paper.

\section{Conclusions} \label{sec:conclusions}

In this work we have analyzed possible sources of bias in the
estimation of the second-order characteristics of the Ly-${\alpha}$
forest. In particular, the metallic and the strong Ly-$\alpha$ lines
have been considered that can hidden part of the lines of interest. 
This may result in a severe bias of both the $K$-function and the two-point
correlation function. Specifically, an erroneous indication of
clustering can happen even in the case of {\it no-interaction} or {\it
complete spatial randomness} in the line pattern or, if a clustering
is really present, a higher degree of clustering than true can be
obtained. In order to handle this problem, we have proposed a computational
efficient method that has been implemented in two different algorithms; one that is
able to provide exact results and the other one that is approximated
but very fast. Its excellent performances have
been validated through numerical simulations and an application to
real data. We have also presented an extension of the method for the estimation of 
the two-point correlation function in the case of two or more lists of lines that have to be analyzed 
at the same time.

\clearpage


\clearpage

\clearpage


\clearpage

\begin{appendix}

\section{An exact method to compute $|W \cap W_r|$} \label{sec:helga}

\begin{figure}
        \resizebox{\hsize}{!}{\includegraphics{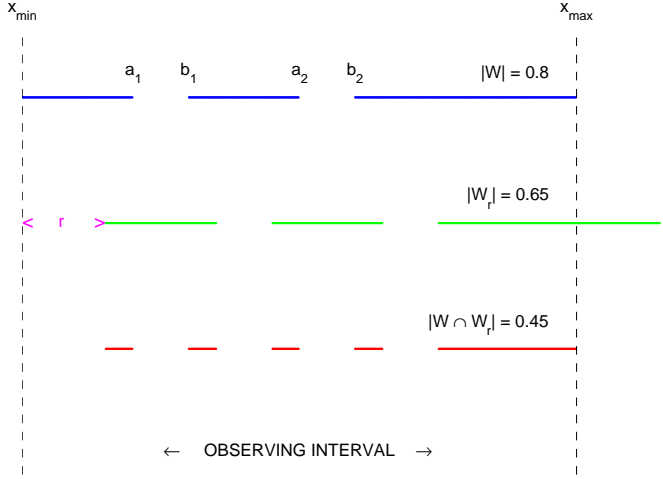}}
        \caption{Example of an {\it observing window} $W$ with two gaps, its shifted version $W_r$ and the
        intersection $W \cap W_r$. The window lengths are given in units of the {\it observing interval} $x_{\rm max}-
        x_{\rm min}$. r=0.15 in the same units.}
        \label{fig:intersect}
\end{figure}

The following presents a pseudo-code that implements an exact algorithm for the computation of
$|W\cap W_r|$ for a window $W$ with gaps. It is assumed that the {\it observing window}
$W$ lies in the interval $[x_{\rm min}, x_{\rm max}]$ and has $n_k$ gaps $[a_k,b_k]$ for
$k=1,...,n_k$ with
\begin{equation}
x_{\rm min}=b_0<a_1<b_1<...<a_{n_k}<b_{n_k}<a_{n_k+1}=x_{\rm max}. \nonumber
\end{equation}
Fig.~\ref{fig:intersect} shows the situation for the case $n_k=2$.

The algorithm considers all combinations of the intervals between
the gaps, determines the lengths of their intersections and sums up.
The wanted length $|W\cap W_r|$ is denoted by $s$.

\noindent
\% input: \\
\% $r$ $\longrightarrow$ interpoint distance \\
\% \\
\\
\% initialization of indices:\\
$s=0;$ \\
$i1=1;$ \\
$i2=1;$ \\
\\
\% fix the shifted interval:\\
$y1= b[i1-1]+r;$\\
$y2=a[i1]+r;$\\
if $y2>x_{\rm max}$ then $y2=x_{\rm max};$ end\\
\% fix the original interval:\\
$z1=b[i2-1];$ \\
$z2=a[i2];$\\
\\
\% compute the length of the intersections \\
while $(i1\leq n_k+1)$ and $(y1<x_{\rm max})$ do\\
\hspace*{1cm} if $z2<y2$ then\\
\hspace*{2cm} if $z2>y1$ then\\
\hspace*{3cm} if $z1>y1$ then \\
\hspace*{4cm} $s=s+z2-z1;$\\
\hspace*{3cm} else \\
\hspace*{4cm} $s=s+z2-y1;$\\
\hspace*{3cm} end\\
\hspace*{3cm} $i2=i2+1;$ \\
\hspace*{3cm} $z1=b[i2-1];$ \\
\hspace*{3cm} $z2=a[i2];$\\
\hspace*{2cm} end\\
\hspace*{1cm} else if $z1<y2$ then\\
\hspace*{3cm} if $z1>y1$ then \\
\hspace*{4cm} $s=s+y2-z1;$\\
\hspace*{3cm} else \\
\hspace*{4cm} $s=s+y2-y1;$\\
\hspace*{3cm} end\\
\hspace*{3cm} $i1=i1+1;$ \\
\hspace*{3cm} $y1=b[i1-1]+r$;\\
\hspace*{3cm} if $i1\leq n_k+1$ then\\
\hspace*{4cm} $y2=a[i1]+r$;\\
\hspace*{4cm} if $y2>x_{\rm max}$ then \\
\hspace*{5cm} $y2=x_{\rm max}$; \\
\hspace*{4cm} end\\
\hspace*{3cm} end\\
\hspace*{2cm} end\\
\hspace*{1cm} end\\
end\\
return $s$

\section{An approximate fast method to compute $|W \cap W_r|$} \label{sec:approximated}

The method presented in the previous section
provides an exact value of $|W \cap W_r|$. If the computing time is
of concern, a more efficient, although approximated
method, is possible. The idea is to consider the {\it observing
window} $W$ as a function $\Wc(x)$ in $\Rb$ defined in the domain
$(-\infty, +\infty)$. When $x \in [x_{\rm min}, x_{\rm max}]$, then
$\Wc(x)=0$ if $x$ belongs to a gap, $\Wc(x)=1$, otherwise. If $x
\notin [x_{\rm min}, x_{\rm max}]$, it is $\Wc(x)=0$.

It is not difficult to realize that, for a fixed $r^*$, $\gamma(r^*)
\equiv |W \cap W_{r^*}|$ is given by 
\begin{equation}
\gamma(r^*) = \int_{-\infty}^{+\infty} \Wc(x) \Wc(x+r^*) dx.
\end{equation}
When $r^*$ is made to change in the interval $(-\infty, +\infty)$, this
  equation provides the autocorrelation function 
of $\Wc(x)$. Hence,
\begin{equation} \label{eq:wapprox}
\gamma(r) = {\rm IFT} \left\{ {\rm FT}[\Wc(x)]  \times ({\rm FT}[\Wc(x)])' \right\}.
\end{equation}
The symbol ``$~{}'~$'' means {\it complex conjugation}, 
whereas ${\rm FT}$ and {\rm IFT} denote the {\it Fourier transform}
and the {\it inverse Fourier transform}, respectively. 

An efficient implementation of Eq.~(\ref{eq:wapprox}) requires the
interval $[x_{\rm min}, x_{\rm max}]$ 
be discretized in a set of $N_b$ bins. In the case of the estimator
$\xih(r)$, this does not represent a true 
limit since it is a function that is computed on spatial bins of
length $2 h$ (see Eq.~(\ref{eq:xi1})). 
The value of $N_b$ to use in the discretization depends on factors as
the extension of the gaps, the 
number of points and the range of distance $r$ of interest (see also below). In any case, this is not 
a critical quantity. In the numerical experiments presented in the previous sections,
$N_b=5000$ is sufficient to obtain results almost indistinguishable
from those obtainable with an exact approach. However, even values of
$N_b$ much larger 
than this one (e.g., $10^5$) determine only a limited increase of the
computational burden. 

If $\wb$ is an array of length $2 N_b$, with its central $N_b$
elements containing function $\Wc(x)$ in the interval 
$x \in [x_{\rm min}, x_{\rm max}]$ sampled on a discrete grid, then
\begin{equation} \label{eq:s1}
\gammab \approx {\rm IDFT} \left\{ {\rm DFT}[\wb] \odot ({\rm DFT}[\wb])' \right\} \times \Delta x.
\end{equation}
Symbol ``$\odot$'' indicates the {\it point-wise multiplication},
$\Delta x$ is the sampling step and ${\rm DFT}$ and {\rm IDFT} denote the fast
{\it discrete Fourier transform} and the fast
{\it inverse discrete Fourier transform}, respectively. The computational benefit of this approach is evident
since the  first $N_b$ elements of $\gammab$ contain the values of $\gamma(r)$ 
for $0 \leq r \leq (x_{\rm max} - x_{\rm min})$.
The fact of using an array $\wb$ of length $2 N_b$ corresponds to a {\it zero-padding} operation to avoid the
aliasing effects typical of the discrete Fourier transform. Array $\gammab$ can be used to
compute the weights $\{ p_{ij} \}$ in Eq.~(\ref{eq:xi1}).

\section{An approximate fast method to compute $\lambda_S(r)$ and $\lambda_V(r)$} \label{sec:lambdas}

As shown by Eqs.~(\ref{eq:k1})-(\ref{eq:xi1}), the estimators
$\Kh(r)$ and $\xih(r)$ require the intensities $\lambda_V(r)$ and
$\lambda_s(r)$. Again, the computation can be done with an exact
algorithm as in Appendix~\ref{sec:helga}. As an alternative, an
approximative but very efficient Fourier based
algorithm is possible that, similarly to
Appendix~\ref{sec:approximated}, requires the discretization of the
interval $[x_{\rm min}, x_{\rm max}]$ in a set of $N_b$ bins.

Once $\gammab$ has been calculated, the computation of $\lambda_S(r)$ is a trivial operation.
In fact, it is sufficient to check whether the elements of the array $\wb$ that correspond to $x_i-r$
and $x_i+r$ are equal to $0$ or to $1$. The situation is more complex for $\lambda_V(r)$ because of
the terms $\theta(r) = \sum_{i=1}^n | W \cap [x_i-r, x_i+r] |$ and $\epsilon(r) \equiv 2
\int_0^r  |W \cap W_t| dt$.

For $\theta(r)$, the basic idea is to identify
the term $[x_i-r, x_i+r]$ as a function $w_{i}^r(x)$ such as
$w_{i}^r(x)=1$ for $x \in [x_i-r, x_i+r]$, $w_{i}^r(x)=0$ otherwise.
The reason is that in this way $w_{i}^r(x)$ can be expressed in the
convolution form
\begin{equation}
w_{i}^r(x) = \int_{-\infty}^{+\infty} I_r(|x|) \delta(x-x_i) dx,
\end{equation}
where $\delta(x)$ is the classic {\it impulse function}. Then, for a given $r^*$, it is possible to
write $\theta(r^*)$ in the form
\begin{equation}
\theta(r^*) = \int_{-\infty}^{+\infty} \Wc(x) \sum_{i=1}^n w_{i}^{r^*}(x) dx.
\end{equation}
Now, a discrete approximation of $\theta(r^*)$ is provided by
\begin{equation} \label{eq:theta}
\theta(r^*) \approx {\rm SUM}\left[ \wb \odot{\rm IDFT} \left\{ {\rm DFT}[\Ib_{r^*}] \odot {\rm DFT}[\xb] \right\}
\right] \times \Delta x.
\end{equation}
${\rm SUM}[\zb]$ is the operator that sums up all the elements of the array $\zb$,
$\xb$ is an $N_b + N_0$ array that in its central $N_b$ elements contains the number of points in each
of the bins of $W$, whereas $\Ib_r$ is a $N_b + N_0$ array containing the discretized version of the function $I_r[.]$
stored in wraparound order with $N_0$ providing the number of zero-pads necessary to avoid the aliasing effect due
to the {\rm DFT}. Here, also the array $\wb$ has size $N_b + N_0$. The term ${\rm DFT}[\Ib_{r^*}]$ has not
necessarily to be calculated numerically since an analytical expression is available \citep[e.g., see][]{bri96}.
In principle, $N_b$ should be chosen large enough that each entry of the array $\xb$ is either $0$ or $1$. However,
from our numerical experiments its comes out that a value of $N_b$ such as $5$-$10$ bins correspond
to the bandwidth $h$ is sufficient to obtain satisfactory results. 

For a given $r^*$, an approximation of $\epsilon(r^*)$ can be obtained from the array $\vec{\gammab}$ by means of
\begin{equation}
\epsilon(r^*) \approx 2 \times {\rm SUM}[\vec{\gammab}(1:i_{r^*})] \times \Delta x.
\end{equation}
$(1:i_{r^*})$ indicates the indices of the elements of the array
to sum up and $i_r$ is the index of the bin corresponding to the distance $r$.

\section{An approximated fast method to compute $\lambda^2 \xih(r)$ and $\lambda^2 \Kh(r)$} \label{sec:functions}

In the previous appendices some efficient numerical methods have been presented to compute
$| W \cap W_r |$ and $\lambda_S(r)$. These are some of the ingredients for the computation of the
function $\xih(r)$ that, however, requires also the total number, $N_h(r)$, of pairs with a 
distance in the range 
$[r - h, r + h]$. The simplest method to obtain this last quantity consists in counting. In the 
one-dimensional case, especially for limited data sets and/or small $r$, this represents the most efficient approach 
(we have used it in our numerical experiments). In the case of large data sets, an alternative method based on a Fourier
approach is possible, that permits to obtain $N_h(r)$ without the necessity to compute the distance between the points
\citep[see also][]{sza05}.

The basic idea is that two points $x_i$ and $x_j$ ($x_i < x_j$) have a distance in the range $[r-h, r+h$]
if $x_j \in [x_i + r -h, x_i+r+h]$. Now, this condition can be expressed in the equivalent form 
\begin{equation}
2 h \int_{-\infty}^{+\infty} k_h(x-x_i-r) \delta(x-x_j) dx = 1.
\end{equation}
A similar expression holds when $x_i > x_j$. Hence, if $r > h$, $N_h(r)$ is given 
\begin{equation}
N_h(r) = 2 h \sum_{i=1}^n \sum_{j=1}^n \int_{-\infty}^{+\infty} k_h(|x-x_i|-r) \delta(x-x_j) dx.
\end{equation}
When $r \leq h$, this number has to be subtracted by the number $n$ of points (to avoid self-pairing).
If discretized, for a fixed $r^*$, this equation takes the form,
\begin{equation}
N_h(r^*) \approx {\rm SUM} \left[ \xb \odot {\rm IDFT} \left\{ {\rm DFT}[\kb_h^{r^*}] \odot {\rm DFT}[\xb] \right\}
\right],
\end{equation}
where $\kb_h^{r^*}$ is the discretized form of the function $2 h k_h(x-r^*)$ stored in wraparound order. At this point,
the computation of $\xih(r)$ should require the calculation of the weights $\{ p_{ij} \}$ that depend
on the quantities $\{ |W \cap W_{x_i-x_j} | \}$. In their turn, these quantities depend on the distances 
$\{ |x_i -x_j| \}$ that are not available. However, if $r \gg h$ (a typical situation of interest),
from Eq.~(\ref{eq:xi1}) it is
\begin{equation} \label{eq:approx1}
\frac{|W|}{|W \cap W_{x_i-x_j}|} k_h(|x_i - x_j|-r) \approx \frac{|W|}{|W \cap W_{r}|} k_h(|x_i - x_j|-r).
\end{equation}
With this approximation the quantity $\lambda^2 \xih(r)$ is given by
\begin{equation}
\lambda^2 \xih(r^*) = \frac{|W|^2 c(r^*)}{2 |W \cap W_{r^*}|} N_h(r^*).
\end{equation}
The denominator of this expression can be obtained from $\gammab$.

In Eq.~(\ref{eq:approx1}), the term ${\rm DFT}[\kb_h^{r^*}]$ can be computed without actually performing 
the {\rm DFT} operation if the array $\kappab_h^0 \equiv {\rm DFT}[\kb_h^{0}]$ is available. In fact, it is
\begin{equation}
\kappab_h^{r^*} = \kappab_h^{0} \odot  {\rm e}^{- \iota 2 \pi i_{r^*} [0, 1, \ldots, N-1]/N},
\end{equation}
with $\iota \equiv \sqrt{-1}$, $i_r$ the index of the bin corresponding to the distance $r$, and $N$ the 
length of the array $\kappab_h^{0}$. 

With arguments similar to those presented above, it is possible to show that, if $N(r^*)$ is the
total number of pairs with distance less or equal to a fixed $r^*$, then
\begin{equation}
N(r^*) \approx {\rm SUM} \left[ \xb \odot {\rm IDFT} \left\{ {\rm DFT}[\Ib_{r^*}] \odot {\rm DFT}[\xb] \right\}
\right] - n,
\end{equation}
and
\begin{equation}
\lambda^2 \Kh(r^*) \approx \frac{|W|^2}{| W \cap W_r|} N(r^*). 
\end{equation}

\end{appendix}


\begin{thebibliography}{}
\bibitem[\protect\citeauthoryear{Ballester et al.}{2000}]{ball00}
Ballester, P., Modigliani, A., Boitquin, O., et al. 2000,
ESO The Messenger, 101, 31

\bibitem[\protect\citeauthoryear{Bergeron et al.}{2004}]{bergeron04}
Bergeron, J., Petitjean, P., Aracil, B., et al. 2004, ESO The
Messenger, 118, 40

\bibitem[\protect\citeauthoryear{Bi \& Davidsen}{1997}]{bi97}
Bi, H.-G., \& Davidsen, A.~F. 1997, \apj, 479, 523

\bibitem[\protect\citeauthoryear{Briggs \& Henson}{1996}]{bri96}
Briggs, W.L., \& Henson, V.E. 1996, Applied Numerical Mathematics,
20, 1

\bibitem[\protect\citeauthoryear{Cen et al.}{1994}]{cen94}
Cen, R., Miralda-Escud\'e, J., Ostriker, J.P., \& Rauch, M. 1994, \apj, 437, L83

\bibitem[\protect\citeauthoryear{Cristiani et al.}{1997}]{cristiani97}
Cristiani, S., D'Odorico, S., D'Odorico, V., et al. 1997, \mnras, 285, 209

\bibitem[\protect\citeauthoryear{Dav\'e et al.}{1997}]{dave97}
Dav\'e, R., Hernquist, L., Weinberg, D.H., \& Katz, N. 1997, \apj, 477, 21

\bibitem[\protect\citeauthoryear{Dekker et al.}{2000}]{dekker}
Dekker, H., D'Odorico, S., Kaufer, A., Delabre, B., \& Kotzlowski, H. 2000,
SPIE, 4008, 534

\bibitem[\protect\citeauthoryear{D'Odorico et al.}{2006}]{dodorico06}
D'Odorico, V., Viel, M., Saitta, F., et al. 2006, \mnras, 372, 1333

\bibitem[Guan (2006)]{gua06} Guan Y. 2006, Statistics and Probability Letters,
{\it submitted}

\bibitem[Hamilton (1993)]{ham93} Hamilton A.J.S. 1993, \apj, 417, 19

\bibitem[\protect\citeauthoryear{Hernquist et al.}{1996}]{hernquist96}
Hernquist, L., Katz, N., Weinberg, D.H., \& Miralda-Escud\'e, J. 1996, \apj,
457, L51

\bibitem[Kerscher (1999)]{ker99} Kerscher, M. 1999, \aap, 343, 333

\bibitem[Kerscher et al. (2000)]{ker00} Kerscher, M., Szapudi, I., \&
  Szalay, A. 2000, \apjl, 535, L13 

\bibitem[\protect\citeauthoryear{Kim, Cristiani \& D'Odorico}{Kim et
    al.}{2001}]{kim01} 
Kim, T.-S., Cristiani, S., \& D'Odorico, S. 2001, \aap, 373, 757

\bibitem[\protect\citeauthoryear{Kim et
    al.}{2002}]{kim02} Kim, T.-S., Carswell, R.F., Cristiani, S., D'Odorico S., \& Giallongo, E.
    2002, \mnras, 335, 555

\bibitem[Landy \& Szalay (1993)]{lan93} Landy ,S.D., \& Szalay,
  A.S. 1993, \apj, 412, 64 

\bibitem[\protect\citeauthoryear{Lu et al.}{1996}]{lu96}
Lu, L., Sargent, W.L.W., Womble, D.S., \& Takada-Hidai,  M. 1996, \apj, 472, 509

\bibitem[Martinez \& Saar (2002)]{mar02} Martinez, V.J., \& Saar,
  E. 2002, Statistics of the Galaxy Distribution 
(Boca Raton: Chapman \& Hall/CRC press)

\bibitem[Mart{\'{\i}}nez et al.(2005)]{mar05}
Mart{\'{\i}}nez, V.J., Starck, J.L., Saar, et al. 2005, \apj, 634, 744

\bibitem[\protect\citeauthoryear{Machacek et al.}{2000}]{machacek00}
Machacek, M. E., Bryan, G. L., Meiksin, A., Anninos, P., Thayer, D., Norman,
M., \& Zhang, Y. 2000, \apj, 532, 118

\bibitem[\protect\citeauthoryear{Miralda-Escud\'e et al.}{1996}]{miralda96}
Miralda-Escud\'e, J., Cen, R., Ostriker, J.P., \& Rauch, M. 1996, \apj, 471, 582

\bibitem[Moller \& Waagepetersen (2004)]{mol04} Moller, J., \& Waagepetersen, R.P. 2004,
Statistical Inference and Simulation for Spatial Point Processes 
(Boca Raton: Chapman \& Hall/CRC press)

\bibitem[Ohser \& M\"uecklich (2000)]{osh00} Ohser, J., \& M\"uecklich, F. 2000,
Statistical Analysis of Microstructures in Materials Science (Chichester: Wiley)

\bibitem[Pons-Border\'ia et al. (1999)]{pon99} Pons-Border\'ia, M.J.,
  Martinez, V.J., Stoyan, D., Stoyan, H., \& Saar, E. 1999, \apj, 523,
  480 

\bibitem[Quashnock \& Sein (1999)]{qua99} Quashnock, J.M., \& Stein,
  M.L. 1999, \apj, 515, 506 

\bibitem[\protect\citeauthoryear{Rauch et al.}{2005}]{rauch05}
Rauch, M., Becker, G.D., Viel, M., Sargent, et al. 2005, \apj, 632, 58


\bibitem[\protect\citeauthoryear{Sargent et al.}{1980}]{sargent80}
Sargent, W.L.W., Young, P.J., Boksenberg, A., \& Tytler, D. 1980, \apjs, 42,
41

\bibitem[Stoyan et al. (1993)]{sto93} Stoyan, D., Bertram, U. \& Wndrock, H. 1993, 
Annals of the Institute of Statistical Mathematics 45, 211.

\bibitem[Stoyan \& Stoyan (2000)]{sto00} Stoyan, D., \& Stoyan
  H. 2000, Scandinavian Journal of Statistics, 27, 641 

\bibitem[Szapudi et al. (2005)]{sza05} Szapudi, I., Pan, J., Prunet, S., \&
budavari, T. 2005, \apj, 631, L1

\bibitem[\protect\citeauthoryear{Theuns, Leonard \&
    Efstathiou}{1998}]{theuns98}
Theuns, T., Leonard, A., \& Efstathiou, G. 1998, \mnras, 297, L49

\bibitem[\protect\citeauthoryear{Zhang, Anninos \& Norman}{Zhang et
    al.}{1995}]{zhang95}
Zhang, Y., Anninos, P., \& Norman, M.L. 1995, \apj, 453, L57

\bibitem[\protect\citeauthoryear{Zhang et al.}{1997}]{zhang97}
Zhang, Y., Anninos, P., Norman, M.L., \& Meiksin, A. 1997, \apj, 485, 496
\end{thebibliography}
\end{document}